\documentclass[aps,pre,floatfix,superscriptaddress,twocolumn,showpacs]{revtex4-1}
\usepackage{amsmath,amssymb,eucal,graphicx}
\usepackage{float}
\usepackage{xcolor, soul,colortbl}

\definecolor{myblue}{HTML}{88AADD}
\definecolor{mygray}{rgb}{0.78,0.78,0.78}

\begin{document}

\title{From anomalous diffusion in polygons to a transport locking relation}

\author{Luisana Claudio-Pachecano}
\thanks{Presently at CENIDET, Apatzingán 212, Palmira, 62490 Cuernavaca, Morelos, México.}
\affiliation{Instituto de Ciencias F\'isicas, UNAM, CP 62210 Cuernavaca Morelos, México}
\author{Hernán Larralde}
\email{hernan@icf.unam.mx}
\affiliation{Instituto de Ciencias F\'isicas, UNAM, CP 62210 Cuernavaca Morelos, México}
\author{Carlos~Mej\'ia-Monasterio}
\email{carlos.mejia@upm.es}
\thanks{ORCID: \href{https://orcid.org/0000-0002-6469-9020}{0000-0002-6469-9020}}
\affiliation{School of Agricultural, Food and Biosystems
Engineering, Technical University of  Madrid, Av. Puerta de Hierro 2, 28040 Madrid, Spain}
\affiliation{Grupo Interdisciplinar de Sistemas Complejos (GISC),  Madrid, Spain}

\begin{abstract}
  We study particle transport in a class of open channels of finite
  length, made of identical cells of connected open polygonal
  billiards with parallel boundaries. In these systems the Mean Square
  Displacement (MSD) grows in time faster than linearly.  We show that
  irrespective of the geometry of these channels, the distribution of
  the first return times decays algebraically with two different
  exponents, separated by a crossover region that is determined by the
  MSD. We find that the distribution of first return times satisfies a
  simple scaling form. In turn, the transmission coefficient, defined
  as the fraction of trajectories that starting at the cell at the
  origin escape the channel through the other boundary, decays
  algebraically with the size of the system, and, as a signature of
  non-recurrent transport, sometimes slower. From these two processes
  we derive a locking relation among the scaling exponents for the
  asymptotic behavior of the MSD, the times of first return to the
  origin and the way transmission decays with the system size, showing
  that these three processes are interdependent. The locking relation
  holds for diffusive processes, as well as for fractional Brownian
  motion with arbitrary Hurst exponent.  We argue that the locking
  relation may be valid for many other transport processes, Markovian
  or not, with finite MSD.
\end{abstract}


\maketitle

\section{Introduction}

The statistics of First Passage Times (FPT) to reach a target is a
fundamental problem in the theory of random walks and stochastic
processes \cite{feller1968,redner2001}. It appears in a wide variety
of problems in physics, chemistry, biology, and economics, like
\emph{e.g.}, the random search of mobile or immobile targets
\cite{mejia2011,metzler2014}, in chemical reactions
\cite{szabo1980,benichou2000,loverdo2008}, the firing of neurons
\cite{gerstein1964,burkitt2006}, protein binding in DNA
\cite{halford2004}, the dynamics of molecular motors
\cite{oshanin2004,newby2010}, animal foraging \cite{ramos2004}, and
the triggering of a stock option \cite{bouchaud2003}, among many
others.

In the presence of multiple targets, a related quantity is the
probability to find a specific target called the hitting or splitting
probability \cite{redner2001}.  In its simplest version, the problem
consists on a diffusing particle in the interval $[0,L]$ with
absorbing boundaries. Moreover, knowing the probabilities that the
walker hits the boundary $0$ or $L$, before a time $t$ is equivalent
to know the survival probability of the walker. An iconic example of
this setting is the classical gambler's ruin problem
\cite{feller1968,redner2001}.  The splitting probabilities have been
studied in variety of phenomena, for diffusive processes
\cite{condamin2007,linn2022}, anomalous diffusion
\cite{scher1975,wiese2019,majumdar2010}, and active particles
\cite{klinger2022,gueneau2024,huang2024}.

In this paper we are interested in studying the transport properties
of particles as they cross a scattering medium of finite length. For a
flux of particles injected at one side of the medium, the transmission
probability is the splitting probability for the particle to exit
through the opposite side before having returned to the side it was
injected from.  This setting was thoroughly analyzed in the seminal
paper of Scher and Montroll on transit times through amorphous
materials \cite{scher1975}, and has been successfully applied to
electronic transport in disordered materials
\cite{bouchaud1990,burioni2010,klinger2022} and chemical transport in
soil \cite{berkowitz2006}, to mention a few.

We consider the evolution of an incoming flow of non-interacting,
point particles moving inside a long, narrow polygonal billiard of
length $L$. Inside this polygonal channel, the particles evolve freely
between specular reflections at the channel flat walls. In the absence
of trapping, the incoming flow is divided in a transmission flow,
corresponding to the particles that being injected through the channel
opening at $0$, exit the channel at $L$; and a reflection flow
composed by the particles that exit the channel through the opening at
$0$.  The main goal of this paper is to determine the statistics of
particle transport in terms of the dynamics of these flows.

The billiard dynamics in polygons is greatly determined by the
geometry, most importantly on whether the polygon is rational
(i.e. the opening angles are rationally related to $\pi$) or not.
Rational polygons have raised great interest recently in the
mathematical areas of algebraic geometry and topology due to recent
advances in moduli spaces in flat surfaces and Teichm\"uller theory
\cite{smillie1999,zorich2005,yoccoz2010,forni2015}. For irrational
polygons not much is known \cite{gutkin1996,bobok2012}.

Billiards dynamics  in polygons are  an example of  pseudointegrable
dynamics:  strictly  speaking,  the  dynamics  is  integrable  as  all
Lyapunov exponents are zero, yet, their  phase space does not have the
simple  topology  of a torus. 

Challenging the common, but not strict, connection between chaotic
dynamics and normal diffusive transport, polygonal channels have been
studied to characterize particle transport
\cite{orchard2024,orchard2021,vollmer2021,jepps2008,sanders2006,jepps2006,alonso2002}.
Particle transport in polygonal channels is strongly and erratically
dependent on the details of their geometry. With parallel boundaries,
transport exhibits strong anomalous diffusion \cite{castiglione1999},
with a superdiffusive MSD and ballistic higher-order moments of the
distribution of the particle displacement $P(x,t)$
\cite{vollmer2021,orchard2021}.  As a consequence, $P(x,t)$ does not
have a simple scaling form.  The ballistic scaling is evident in the
tails of $P(x,t)$ in which several ballistic fronts, with different
speeds of propagation, can be observed. In \cite{orchard2024} the
speeds of the ballistic fronts have been analytically derived for a
rational channel with parallel walls. With non-parallel boundaries,
diffusion and even subdiffusion has been observed
\cite{sanders2006,jepps2006,jepps2008,alonso2002}.

Borrowing ideas from stochastic dynamics, we will idealize the flow of
particles as individual deterministic walkers, and randomness entering
strictly from their initial conditions. Following the program of
\cite{lichtenberg1992} (see also \cite{orchard2021}), our approach is
to study how systems with deterministic dynamics may exhibit features
typically associated with stochastic dynamics.

We focus first on the statistics of the first return times
$P_0(\tau)$, where $\tau$ is the time it takes a particle that having
entered the channel at $0$ to exit the channel through the same
opening. The ensemble of these trajectories yields the ratio of
reflection. The distribution $P_0(\tau)$ can determine whether the
random walk is recurrent or transient \cite{redner2001}, and relates
other important problems such as survival and persistence
\cite{bray2013}.  In phase space, first return times relate to mixing
properties of deterministic dynamical systems \cite{haydn1999}.

Markovian random walks with independent symmetric steps form an
important class that satisfies the Sparre-Andersen theorem
$P_0(\tau)\sim\tau^{-3/2}$, irrespective of the distribution of step
lengths \cite{sparre1954}. Our polygonal channels are not in this
class. Moreover, we find that the distribution of first return times
decays algebraically with two different exponents, and the
intermediate region separating these scaling regimes is determined by
the Mean Square Displacement (MSD). More importantly, we show that if
the MSD is known, $P_0(\tau)$ can be written in a simple scaling form.
From this we derive an expression for the probability of transmission
that leads to a locking relation between different scaling exponents:
the exponent with which the MSD scales in time, the exponent for the
decay of $P_0(\tau)$, and the exponent with which the transmission
coefficient decays with the size of the system.  Furthermore, we show
that the locking relation holds exactly for diffusive transport and
for fractional Brownian motion describing anomalous transport
\cite{mandelbrot1968}. We speculate that this locking relation may
hold for many other transport process possessing a finite MSD.

The paper is organized as follows: In section~\ref{sec:model} we
define the polygonal billiard channels of parallel walls that will be
used in the rest of the paper. In section~\ref{sec:msd} we review the
properties of particle transport in these polygonal channels. In
section~\ref{sec:escape} we discuss the distribution of first return
times for different geometries of the polygonal channel and show that
it is scale invariant. The locking relation is derived and discussed
in section~\ref{sec:relation}, and in section~\ref{sec:fBM} is shown
that it holds for fBM. In section~\ref{sec:concl} we summarize our
results.

\begin{figure}[!t]
\centering
   \includegraphics[width=0.42\textwidth]{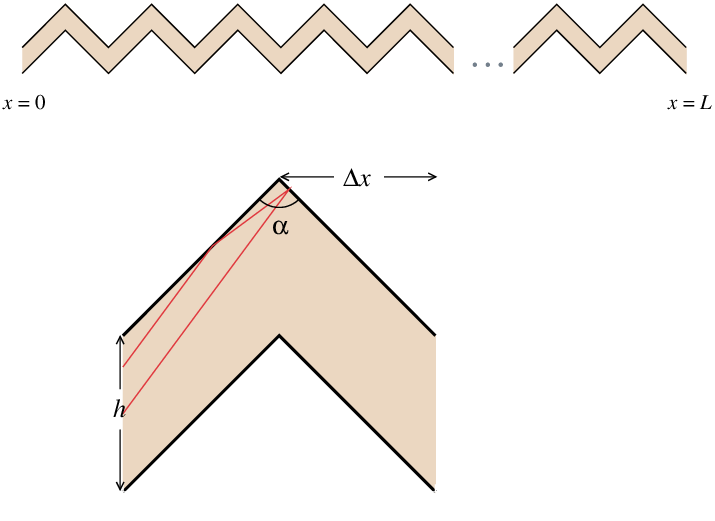}
   \caption{The geometry of the elementary cell is determined by its
     semi-length $\Delta x$, opening angle $\alpha$ and width $h$. The
     upper figure shows a channel of $L$ cells. In red an example of a
     trajectory that returns to the left opening after two
     collisions.}
    \label{fig:model}
\end{figure}

\section{Model}
\label{sec:model}
 
We consider a quasi one-dimensional channel made of $L$ identical
copies of open polygonal billiards.  The elementary polygonal cell
consists of two identical polygonal boundaries, formed by two linear
segments forming an angle $\alpha \in (0,\pi)$, separated by a
distance $d$ (see Fig.~\ref{fig:model}). The extension of the billiard
cell along the transport direction is set to unity, so that
$\Delta x = 1/2$, and the channel extends from $x=0$ at the left to
$x=L$ at the right.

Point particles move freely inside the polygonal cell at constant
speed, and collide elastically with its boundaries. When a particle
reaches any of the cell's openings, it leaves that cell and continues
its motion in the next cell to the left or to the right
accordingly. Note that for $h \le \Delta x \cot(\alpha/2)$ the length
of the trajectories between two consecutive collisions is always
bounded, a condition known as finite horizon.

To study the dispersion properties of an ensemble of particles inside
the polygonal channel we consider the following experiment: At time
$t=0$ each particle is placed at the left opening of the channel, with
coordinates $x=0$ and $y$ uniformly distributed in $(0,h)$.  All the
particles have unit speed with initial direction $\theta$ uniformly
distributed in $(-\pi/2,\pi/2)$. Each particle travels within the
channel until it exits through the left or right openings of the
channel, at a time $t=\tau$. If the particle exits through the initial
opening at the left of the channel, then we say the particle was
reflected, otherwise, we say the particle was transmitted.

Clearly,  the  dynamics  of  each particle  are  fully  deterministic,
meaning  that the  time $\tau$  and  whether the  particle returns  or
crosses the channel  are completely determined by the  geometry of the
cell  and its  initial  condition.  However, the  exact  fate of  each
particle is generally very  hard to predict. In Ref.\cite{orchard2024}
some of us derived analytically the  scattering map of the single cell
of the rational polygon with  $\alpha=\pi/2$, showing that already for
the single cell the structure of the scattering is not simple. Here we
will show that the flow of particles can be thought of as some kind of
random process for many purposes.

\section{Anomalous anomalous transport}
\label{sec:msd}

The statistics of the particle displacement in polygonal channels has
been considered in Refs.~\cite{alonso2004,sanders2006,jepps2006} and
more recently in Refs.~\cite{orchard2021,vollmer2021,orchard2024}.
Transport through polygonal channels exhibits a rich behavior, from
subdiffusion to superdiffusion, strongly depending on the channel's
geometry. Moreover, transport in these systems is characterized by
several unique properties, making it "anomalous" among anomalous
diffusive processes. In this section we review some of these
properties.

\begin{figure}[!t]
\centering
   \includegraphics[width=0.45\textwidth]{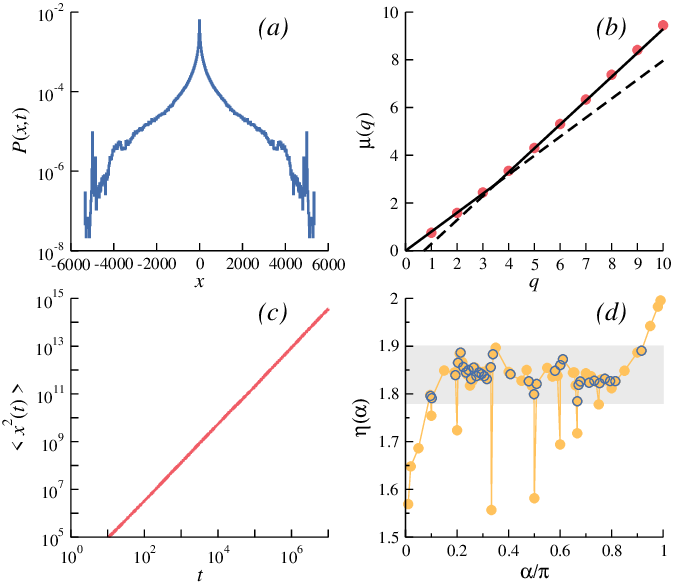}
   \caption{Characteristic anomalous transport of polygonal billiards,
     with opening angle $\alpha=\pi/2$ and critical
     horizon. (\emph{a}) Probability density function of the particles
     displacement $P(x,t)$ measured at a time $t=10^4$. (\emph{b})
     Spectrum of the moments of displacement $\mu(q)$ exhibiting
     biscaling (red circles). The straight lines stand for
     Eq.~\ref{eq:SAD} with the corresponding exponent $\eta=1.5929$
     and $q^\star= 3.5$. (\emph{c}) MSD scaling as
     $\langle x^2(t) \rangle \sim t^{1.5929}$. (\emph{d}) Scaling
     exponent of the MSD $\eta$ as a function of the opening angle
     $\alpha$ (yellow circles). Open blue circles indicate irrational
     angles of the form $\alpha=\xi\pi/p$ (where
     $\xi=\varphi - \frac{3}{5}\approx 1.01803$, $\varphi$ is the
     golden ratio, see text). For these "irrational" cells, the MSD
     scales superdiffusively with exponent
     $1.8 \lesssim \eta \lesssim 1.9$, as indicated by the shaded
     region.}
    \label{fig:transport}
\end{figure}

From an initial uniform flow of particles in an infinite channel,
transport can be characterized in terms of the probability
distribution of the particle displacement $P(x,t)dx$, which is the
fraction of particles found at a distance between $x$ and $x+dx$ from
its initial position at time $t$.  For the present polygonal billiard
channels, the probability density $P(x,t)$ is qualitatively described
in the bulk by a pronounced peak following a stretched exponential
law, and asymptotic tails with ballistic fronts moving out at
different speeds.  Fig.~\ref{fig:transport}(\emph{a}) shows an example
of $P(x,t)$ for the polygonal channel with $\alpha=\pi/2$ and critical
horizon (that is $h = \Delta x \cot(\alpha/2)$). For this channel, the
bulk of the distribution is described by
$\exp\{-k\left(x^{1/4}\right)\}$, with $k$ a positive constant.

A consequence of these particularly anomalous statistics is that small
and large fluctuations are characterized by different space-time
scaling, characteristic of Strong Anomalous Diffusion (SAD)
\cite{castiglione1999}. For the present polygonal channels, the
moments of the displacement follow a "biscaling" law given by
\begin{equation} \label{eq:SAD}
\langle |x|^q \rangle \sim \left\{
\begin{array}{ll}
t^{\frac{\eta}{2} \ q} \ , & \mathrm{for} \ \  q < q^\star \\
\\
t^{q+q^\star(\frac{\eta}{2}-1)} \ , & \mathrm{for} \ \  q > q^\star
\end{array}
\right. \ ,
\end{equation}
where $\eta$ is the scaling exponent of the MSD
$\langle x^2(t) \rangle \sim t^\eta$, and $q^\star > 2$.
Fig.~\ref{fig:transport}(\emph{b}) shows the piecewise linear spectrum
of moments of the displacement, defined as
$\mu(q) = \lim_{t\rightarrow\infty} \log(\langle |x|^q
\rangle)/\log(t)$ for the same polygonal channel as in
Fig.~\ref{fig:transport}(\emph{a}).

Since all ballistic fronts in the density $P(x,t)$ decay rapidly, in
the long time limit, transport is well characterized by the
MSD. However, it has been shown that the MSD depends strongly on the
geometry of the polygonal cell \cite{sanders2006,orchard2021}, even
motivating novel definitions of complexity \cite{jepps2006}.  When
upper and lower boundaries are simple translations of each other (as
in Fig.~\ref{fig:model}), we find that $\eta>1$, yielding
superdiffusive transport for all angles
$\alpha$. Fig.~\ref{fig:transport}\emph{c} shows the MSD for the
polygonal channel with $\alpha=\pi/2$ and critical horizon. From a fit
to power-law we obtain $\eta=1.59$. For other angles the MSD scales
superdiffusively but with an exponent $\eta$ that fluctuates strongly
with the angle, as shown in
Fig.~\ref{fig:transport}\emph{d}. Generically, we find that transport
is strongly superdiffusive, with $1.8<\eta<1.9$, for
$\alpha\lesssim 0.9\pi$, as indicated by the shaded region in
Fig.~\ref{fig:transport}\emph{d}. All irrational polygonal channels
(open blue circles in the figure), that is channels for which
$\alpha/\pi$ is irrational, are generic in this sense. In contrast,
some particular rational channels $\alpha=m\pi/n$, with $m, n$
integers (solid yellow circles), \emph{e.g.}
$\alpha=\{\pi/3, \pi/2, 2\pi/3, 3\pi/4\}$, are characterized by
distinctively lower values of $\eta$. The same is observed for
rational channels with $\alpha/\pi\ll 1$. For
$\alpha/\pi\rightarrow 1$, corresponding to the limit of flat
channels, $\eta\rightarrow 2$ as could be expected.  Here and in what
follows we probe irrational values of $\alpha$ that are close to
rationally related values by multiplying $\alpha$ by
$\xi = \varphi - \frac{3}{5}\approx 1.01803$, where
$\varphi = \left(1+\sqrt{5}\right)/2$ is the golden ratio.

\section{The escape problem}
\label{sec:escape}

We consider now finite polygonal channels made of $L$ cells so that a
particle is inside the channel if the $x$-coordinate of its position
is between $0\le x \le L$ (see Fig.~\ref{fig:model}). In this section
we are interested in characterizing the process by which a particle
initially inside the channel exits it for the first time. This is
customarily solved in terms of the survival probability $S(\tau|x_0)$
defined as the probability that a particle, initially located inside
the channel at $x_0$, is still inside the channel at time $\tau$. The
escape probability is then defined as
\begin{equation} \label{eq:W}
  W(\tau|x_0) = 1 - S(\tau|x_0) \ ,
\end{equation}
and the probability density of the escape time $P_{[0,L]}(\tau|x_0)$,
also called residence time, is given by
\begin{equation} \label{eq:escape}
P_{[0,L]}(\tau|x_0) =  \frac{\partial W(\tau|x_0)}{\partial t} \ .
\end{equation}
Having the possibility of exiting the channel through the borders at
$x=0$ or $x=L$, the escape probability is the result of two
contributions
\begin{equation} \label{eq:splitting}
P_{[0,L]}(\tau|x_0) = P_{[\underline{0},L]}(\tau|x_0)  +
P_{[0,\underline{L}]}(\tau|x_0) \ ,
\end{equation}
where $P_{[\underline{0},L]}(\tau|x_0)$ is the probability that a
particle, initially at $0\le x_0\le L$, exits the channel by crossing
the opening at $x=0$ at time $\tau$ (before crossing the opening at
$x=L$), and $P_{[0,\underline{L}]}(\tau|x_0)$ is the probability that
the particle crosses the channel opening at $x=L$ at time
$\tau$. These two contributions are called hitting or splitting
probabilities \cite{vanKampen1992,redner2001,klinger2022}.

Considering a distribution of initial conditions at the left opening
of the channel, with $x=0$, $y$ uniformly distributed in the interval
$(0,h)$, and with initial direction $\theta_0\in(-\pi/2,\pi/2)$
uniformly as well, the first contribution
$P_0(\tau) = P_{[\underline{0},L]}(\tau|x_0)$ is the probability
density of the first return times or \emph{reflection} times, while
the second contribution $P_L(\tau) = P_{[\underline{0},L]}(\tau|x_0)$
is the probability density of the \emph{transmission} times.

When the underlying dynamics of transport is a Markovian random walk,
the statistics of the first return times is
given by the Sparre-Andersen theorem \cite{sparre1954}, stating that
the probability that a random walk process that starting at the origin
enters the positive (or negative) semi-axis for the first time after
$n$ steps decays, in the limit of large $n$ as $n^{-3/2}$,
independently of the details of the jump length distribution, provided
that it is symmetric.


Even though the dynamics in our polygonal channels appear to satisfy
the conditions of the theorem, they do not belong to the class of
processes described by the Sparre-Andersen theorem, presumably due to
long-range spatial and temporal correlations of the deterministic
dynamics, although extensions of the theorem to processes with
correlations exist (see \emph{e.g.} \cite{artuso2014}).

We have numerically computed the probability density of the first
return times to the origin $P_0(\tau)$ for several polygonal channels.
At very short times $\tau\lesssim 1$, the probability density of the
first return times $P_0(\tau)$ exhibits strong oscillations. This
short time regime is the result of trajectories entering the channel
and being reflected after a single collision with the polygonal
cell. Since these trajectories do not explore the channel, the
properties of $P_0(\tau)$ at these time scales are neither universal
nor interesting.

\begin{figure}[!t]
\centering
   \includegraphics[width=0.5\textwidth]{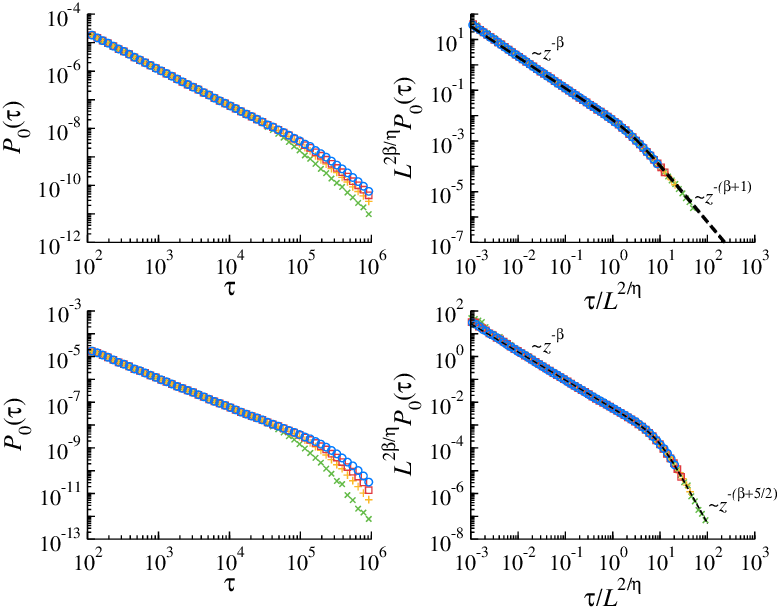}
   \caption{Probability density of the first return times $P_0(\tau)$
     for a channel with $\alpha=\pi/2$ (top), and $\alpha=\xi\pi/2$
     (bottom), with length $L=2500$ (green crosses), $L=5000$ (yellow
     pluses), $L=7500$ (red squares), and $L=10000$ (blue circles).
     The right column shows the collapse of all data points as
     suggested by Eq.~\ref{eq:collapse}. The dashed curves correspond
     to Eq.~\ref{eq:ratfit} for $A=0.018$, $c=2.5$ and $\zeta=1$ (top),
     and to Eq.~\ref{eq:irrfit} for $A=1.5$, $c=9.5$ and $\zeta=5/2$
     (bottom). In both cases the width of the channel is set to
     correspond to critical horizon.}
    \label{fig:collapse}
\end{figure}

At later times the density of first return times decays algebraically
as $P_0(\tau)\sim\tau^{-\beta}$ with a scaling exponent $1<\beta<3/2$
for all $\alpha$, thus confirming that the dynamics of our model
system does not belong to the class of Sparre-Andersen. This decay is
robust and extends for several decades. Interestingly, at longer times
$\tau>\tau^\star$, a second faster algebraic decay appears
$P_0(\tau)\sim\tau^{-\gamma}$. Two examples of $P_0(\tau)$ are shown
in Fig.~\ref{fig:collapse} for a rational polygon with $\alpha=\pi/2$
(upper left), and for an irrational polygon (bottom left) with
$\alpha=\xi\pi/2$.  In both cases $P_0(\tau)$ is shown for different
channel sizes: $L=2500$ (green crosses), $L=5000$ (yellow pluses),
$L=7500$ (red squares), and $L=10000$ (blue circles). Noting that the
crossover time $\tau^\star$ scales with the system size, the
statistics of the first return times can be cast as
\begin{equation} \label{eq:FRT}
P_0(\tau) \sim \left\{
\begin{array}{ll}
\tau^{-\beta} \ , & \mathrm{for} \ \ \tau < \tau^\star \\
\\
L^b\tau^{-\gamma} \ , & \mathrm{for} \ \ \tau < \tau^\star
\end{array}
\right. \ . 
\end{equation}
where $\tau^\star$, $\beta$, $\gamma$ and $b$ are all functions of the
angle $\alpha$. In appendix~\ref{app} we show $P_0(\tau)$ for other
several values of $\alpha$. At sufficiently large times, an
exponential decay due to the finite length of the system could be
expected, but we did not observe it in our simulations.

Using continuity at the crossover $\tau=\tau^\star$, and assuming that
$\tau^\star$ scales with system size as $\tau^\star\sim L^{2/\eta}$
the exponents characterizing $P_0(\tau)$ must satisfy, for arbitrary
$\alpha$ and $L$, that $L^{-2\beta/\eta} = L^{b-2\gamma/\eta}$, from
where we obtain $b=2(\gamma-\beta)/\eta$.  Using this relation between
the exponents the probability of first return times can be written as
\begin{equation}
  P_0(\tau) \sim L^{-2\beta/\eta}\left\{
\begin{array}{ll}
\left(\frac{\tau}{L^{2/\eta}}\right)^{-\beta} \ , & \mathrm{for} \ \
                                                   \frac{\tau}{L^{2/\eta}}
  \ll 1\\
\\
\left(\frac{\tau}{L^{2/\eta}}\right)^{-\gamma} \ , & \mathrm{for} \ \
                                                    \frac{\tau}{L^{2/\eta}}
  \gg 1
\end{array}
\right. \ ,
\end{equation}
from where we obtain that the statistics of the first return times
obey the following scaling
\begin{equation}\label{eq:collapse}
  L^{2\beta/\eta} P_0(\tau) \approx \mathcal{G}\left(
    \frac{\tau}{L^{2/\eta}}\right) \ .
\end{equation}
where
\begin{equation*}
\mathcal{G}(z)\sim
\begin{cases}
z^{-\beta}, & z\ll 1, \cr
\\
z^{-\gamma} & z\gg 1 \cr
\end{cases}.
\end{equation*}
This scaling function will be discussed further on.
Equation~\ref{eq:collapse} constitutes a central result of this paper.

To test Eq.~\ref{eq:collapse} we have numerically computed for several
polygonal channels the MSD and the distribution of the first return
times and from fits to power-law, extracted the exponents $\eta$ and
$\beta$ respectively.  Some of these values are listed in
table~\ref{tab:exponents}.  In the right column of
Fig.~\ref{fig:collapse} we show $P_0(\tau)$ with the scaling of
Eq.~\ref{eq:collapse}.  The remarkably good collapse of the
distributions for the different channel lengths verifies the validity
of Eq.~\ref{eq:collapse}.  This is also verified for all the
investigated channels (see Appendix~\ref{app}).

\begin{table}[!h]
     \centering
\begin{tabular}{|c|c|c|c|c|}
   \hline
 $\alpha$ & $\eta$ & $\beta$ & $\zeta$ & $\nu$  \\
  \hline
  $  \pi/5$ & $1.7309$ &  $1.2193$ & $3/2$ & $0.2533$\\
  $  \pi/4$ & $1.8324$ &  $1.2593$ & $1$ & $0.2830$ \\
  $  \pi/3$ & $1.5566$ &  $1.2830$ & $1$ & $0.3631$ \\
  $  \pi/2$ & $1.5929$ &  $1.2559$ & $1$ & $0.3213$ \\
  $2\pi/3$ & $1.7175$ &  $1.1794$ & $3/2$ & $0.2008$ \\
  $3\pi/4$ & $1.7780$ &  $1.1606$ & $3/2$ & $0.1807$ \\
  \hline
  $  \xi \pi/5$ & $1.865$ & $1.252$ & $5/2$ & $0.271$ \\
  $  \xi \pi/4$ &  $1.831$ & $1.243$ & $5/2$ & $0.265$ \\
  $  \xi \pi/3$ &  $1.883$ & $1.221$ & $5/2$ & $0.235$ \\
  $  \xi \pi/2$ &  $1.821$ & $1.255$ & $5/2$ & $0.280$ \\
  $\xi 2\pi/3$ &  $1.826$ & $1.213$ & $5/2$ & $0.233$ \\
  $\xi 3\pi/4$ &  $1.828$ & $1.183$ & $5/2$ & $0.200$ \\
  \hline
  BM & $1$ & $3/2$ & $ $ & $1$ \\
  fBM & $2H$ & $2-H$ & $ $ & $(1-H)/H$ \\
\hline
\end{tabular}
\caption{Scaling exponent of the MSD $\eta$ and of the algebraic
  decays of $P_0(\tau)$, $\beta$ and $\gamma=\beta+\zeta$ obtained
  numerically from fits to a power-law for different polygonal
  channels. The last column reports the scaling exponent for the decay
  of the transmission coefficient with the system size $\nu$ as
  obtained from Eq.~\ref{eq:relation}. The last rows list the values
  of $\eta$, $\beta$ and $\nu$ for Brownian motion and fractional
  Brownian motion.
  \label{tab:exponents}}
\end{table}

The validity of Eq.~\ref{eq:collapse} also supports the scaling of the
crossover time $\tau^\star=L^{2/\eta}$, corresponding to the time it
takes a particle to effectively explore a system of size $L$.  At
times larger than $\tau^\star$ particles may escape through the right
opening of the channel and therefore, contribute less to the
statistics of the return times. Therefore, it is natural to observe
that the the density $P_0(\tau)$ is convex, namely that
$\gamma>\beta$. However, we have no explanation for the existence of
the second algebraic decay, nor a theory for the value of the exponent
$\gamma$. Note, for instance, that for diffusive dynamics, $P_0(\tau)$
decays exponentially at times larger than $L^2$.

To describe this second algebraic decay we have tested a fitting
function of the rescaled time $z=\tau/L^{2/\eta}$, that interpolates
between both algebraic decays, given by
\begin{equation} \label{eq:ratfit}
  \mathcal{G}_R\left(z\right) =
  \frac{A z^{-\beta}}{\left(c^{2\zeta} + z^{2\zeta}\right)^{1/2}} \ ,
\end{equation}
such that $\mathcal{G}_R\sim z^{-\beta}$ for $z\ll c$, and
$\mathcal{G}_R\sim \tau^{-(\beta+\zeta)}$ for $z\gg c$, with
$\zeta=\gamma-\beta$. The coefficient $c$, that is related to the
generalized diffusion constant, controls the position of the crossover
which in all cases is $c=\mathcal{O}(1)$, and $A$ is determined by
normalization.

It turns out that with the exception of a few particular polygons,
$\mathcal{G}_R\left(z\right)$ fits remarkably well the universal
behaviour of $P_0(\tau)$, including the crossover, as shown by the
dashed curve in the upper right panel of Fig.~\ref{fig:collapse}, for
the rational polygonal channel with $\alpha=\pi/2$.

Furthermore, we have found that for irrational polygons, a better
fitting function is given by
\begin{equation} \label{eq:irrfit}
  \mathcal{G}_I\left(z\right) =
  \frac{A z^{-\beta}}{c^\zeta + z^\zeta} \ ,
\end{equation}
where the parameters have the same interpretation as in
Eq.\ref{eq:ratfit}. Equation \ref{eq:irrfit} has the same asymptotics
as Eq.~\ref{eq:ratfit}, but adjusts the crossover better (see
\emph{e.g.}  the lower right panel of Fig.~\ref{fig:collapse}, for
$\alpha=\xi\pi/2$).

The good agreement of these fitting functions to the scaled
$P_0(\tau)$ suggests that the scaling exponent of the second algebraic
decay $\sim\tau^{-\gamma}$, is not independent of the exponent $\beta$
of the earlier decay, but it is given by $\gamma = \beta + \zeta$,
with $\zeta$ an integer or half-integer that depends on $\alpha$.

\section{Transport locking relation}
\label{sec:relation}

Integration of the probability density of the first return times to
the origin yields the reflection coefficient
\begin{equation} \label{eq:R}
    R(\alpha,L) = \int_0^\infty P_0(\tau) ~\mathrm{d}\tau \ ,
\end{equation}
which in general depends on the geometry and the channel length. The
reflection coefficient corresponds to the fraction of particles that
entering the channel eventually exit the channel through the same
side.

Using Eq.(\ref{eq:FRT}), Eq.(\ref{eq:R}) is given by
\begin{equation} \label{eq:R-2}
    R(\alpha, L)\approx k_0+k_1\int\limits_a^{\tau^\star}\tau^{-\beta}
    ~\mathrm{d}\tau + k_2 L^b\int\limits_{\tau^\star}^\infty
    \tau^{-\gamma} ~\mathrm{d}\tau \ ,
\end{equation}
where $k_0$ is the contribution from the trajectories that bounce out
immediately, $a$ is a time of order unity, $k_1$ and $k_2$ are
constants. Performing the integrals and recalling that
$b=2(\gamma-\beta)/\eta$ gives
\begin{equation} \label{eq:R-3}
    R(\alpha, L)=\mathcal{Q}-\mathcal{K}~L^{-2(\beta-1)/\eta}
\end{equation}
where $\mathcal{Q}$ and $\mathcal{K}$ are positive constants that do
not depend on $L$ (but can depend on $\alpha$).

Assuming that in the limit $L\to\infty$ \emph{all} the trajectories
return to the entrance side, up to a set of measure zero, then
$\mathcal{Q}=1$. In this case we say that transport is recurrent.
Furthermore, integrating Eq.(\ref{eq:splitting}) over $\tau$, the
transmission coefficient $T(\alpha,L)$ which is the fraction of
particles that escape through the right boundary of the channel at
$x=L$, is simply $T(\alpha,L) = 1 - R(\alpha,L)$, and from
Eq.(\ref{eq:R-3}) it implies that if transport is recurrent, then
\begin{equation} \label{eq:T}
T(\alpha,L)\sim \mathcal{K}~L^{-2(\beta-1)/\eta} \ .
\end{equation}
This constitutes our second main result: If for a given polygonal
channel of angle $\alpha$, transport is recurrent, then the
transmission coefficient decays with the system size algebraically as
$T(\alpha,L) \sim L^{-\nu}$, with a scaling exponent given by
\begin{equation} \label{eq:relation}
 \nu = \frac{2\left(\beta - 1\right)}{\eta} \ .
\end{equation}

Equation (\ref{eq:relation}) implies that the scaling of the MSD, the
decay of the transmission coefficient with the system size, and the
statistics of the first return times to the origin are interdependent
processes.

However, polygonal billiards are often non-recurrent, meaning that the
probability that a trajectory never returns to the origin is non zero
\cite{conze2012}. Proving non-recurrence is not an easy task, but one
of its consequences is that the survival probability $S(\tau)$,
i.e. the probability that a particle is still in the channel at time
$\tau$ in the limit of infinitely long channel, decays as
\begin{equation} \label{eq:surv}
  S(\tau) \approx C_0 + C \tau^{-\theta} \ ,
\end{equation}
where $C_0>0$ is the fraction of trajectories that never return to the
origin, and $C>0$ is a normalization constant. For recurrent processes
$C_0=0$, and the exponent $\theta>0$ is known as the persistence
exponent \cite{bray2013}.  For non-recurrent processes,
Eq.(\ref{eq:surv}) describes a survival probability that does not
vanish. In such cases $\mathcal{Q}<1$ in eq.(\ref{eq:R-3}), and the
transmission coefficient becomes
\begin{equation} \label{eq:T-2}
T(\alpha,L) \sim 1-\mathcal{Q}+\mathcal{K}~L^{-2(\beta-1)/\eta} \ .
\end{equation}
We note that when plotted, this expression might look like a pure
power law with a different exponent at short times if $1-\mathcal{Q}$
is small.

\begin{figure}[!t]
\centering
   \includegraphics[width=0.45\textwidth]{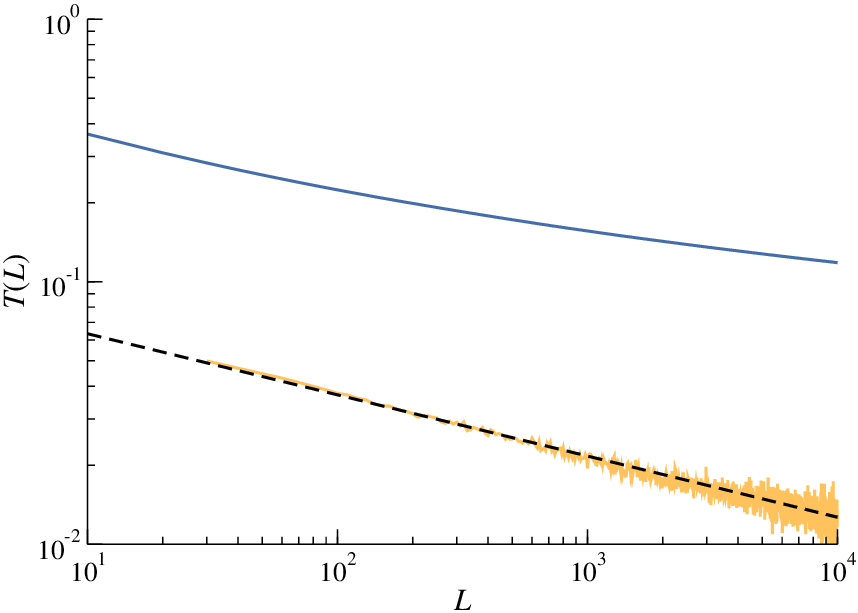}
   \caption{Transmission as a function of the system size $T(L)$ for a
     channel with $\alpha=\xi 2\pi/3$ and critical horizon (dark
     blue). The yellow curve corresponds $-L T^\prime(L)$ and the
     dashed curve to $\sim L^{-\nu}$ with $\nu=0.235$.}
 \label{fig:transmission}
\end{figure}

Fig.~\ref{fig:transmission} shows the transmission coefficient as a
function of the channel length $T(L)$ for a channel with
$\alpha=\xi 2\pi/3$ and critical horizon (dark blue curve). The slower
than algebraic decay implies that for this channel, transport is
non-recurrent. One way to estimate the exponent $\nu$ numerically is
to calculate the derivative with respect to $L$ of $T(\alpha,L)$ and
plot $-LT^\prime(L)$. This corresponds to the yellow curve in
Fig.~\ref{fig:transmission}. Comparing the decay $-LT^\prime(L)$ with
$\sim L^{-2(\beta-1)/\eta}$ (dashed black curve), confirms the
validity of the Eq.(\ref{eq:relation}).

In polygonal billiards non-recurrence is often the result of
trajectories in the neighborhood of non-escaping periodic orbits
\cite{dettmann2010}. In the present polygonal channels, it is likely
to be the result of a finite measure of trajectories in the ballistic
branch yielding an extremely slow decay of correlations.

The exponent relation in Eq.(\ref{eq:relation}) should hold
irrespective of the geometry of the polygonal cell. To show that this
is indeed the case, we have numerically obtained the statistics of the
displacement and of the first return times, as well as the
transmission coefficient $T(L)$ for channels of size $L\in[10,10000]$,
for different polygonal channels with rational and irrational angles
$\alpha$. The scaling exponents $\eta$, $\beta$ and $\nu$ are then
obtained from fits to a power-law as follows: The MSD was obtained for
times up to $10^7$ and averaged over $\approx 5000$ realizations;
while the exponent $\eta$ was obtained from a fit to a power-law in
the time interval $[10^4,10^7]$.  To obtain $\beta$ we first scaled
$P_0(\tau)$ computed for a channel of length $L=10000$ cells, as in
Eq.(\ref{eq:collapse}). Then $\beta$ was obtained from a fit to a
power-law in the time interval
$-2.6 < \ln\left(\frac{\tau}{L^{2/\eta(\alpha)}}\right) < -1.2$.
After scaling, the decay $\sim \tau^{-\beta}$ appears clearly for all
values of $\alpha$. Finally, to deal with possible non-recurrent
transport, we first computed numerically the derivative of the
transmission coefficient; then $\nu$ is obtained from a fit to a
power-law of the function $-LT^\prime(L)$ in the interval
$100<L<1900$. We chose this interval to avoid fitting the noisy signal
that is obtained by computing the numerical derivative.

\begin{figure}[!t]
\centering
   \includegraphics[width=0.47\textwidth]{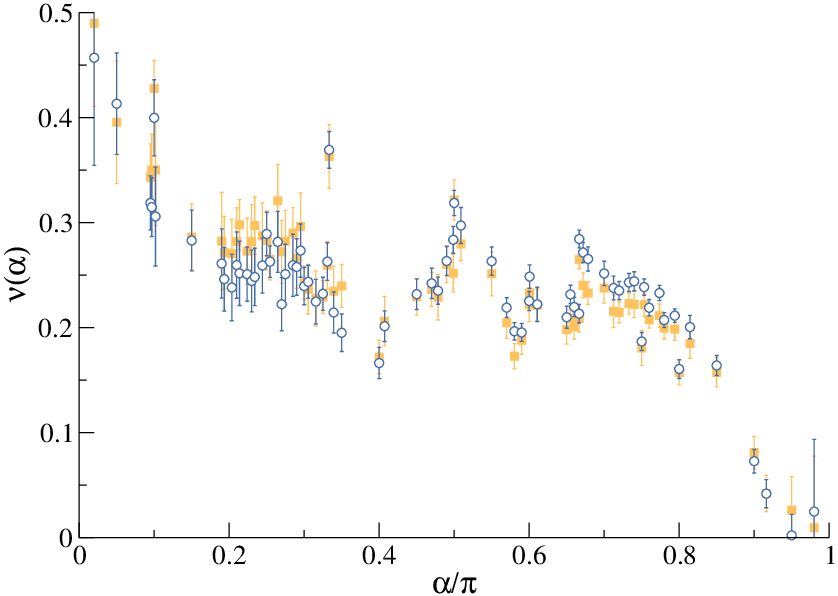}
   \caption{Transmission scaling exponent $\nu$ as a function of
     $\alpha$ and critical horizon (open blue circles), and its
     estimation from Eq.~\ref{eq:relation} (yellow solid circles). The
     error bars were obtain after bootstrapping.}
    \label{fig:relation}
\end{figure}

Fig.\ref{fig:relation}, compares the scaling exponent $\nu(\alpha)$
(open blue circles), obtained as explained above, with
Eq.(\ref{eq:relation}) (full yellow squares), estimated from the
numerical results for $\eta(\alpha)$ and $\beta(\alpha)$.  Error bars
are the estimated standard deviation obtained by means of
bootstrapping. The agreement is fairly good, particularly considering
the noise that arises from computing the numerical derivative of
$T(L)$.  Over a large interval of angles, the values of $\nu$ spread
around $0.25$, and converge to $\nu=0$ as $\alpha\rightarrow\pi$. The
latter is expected since for $\alpha=\pi$ the channel becomes straight
and transmission does not scale with the system size.

It is worthwhile noting that for some polygons, \emph{e.g} for
$\alpha=\pi/4$ (see Fig.~\ref{ap:0.2500} in the appendix), the
derivative of the transmission $T(L)$ still decays slower than
algebraically.  While we have no explanation for this behavior, we
speculate that additional corrections to $T(L)$ may be due to super
slow modes in the decay of correlations for certain geometries.
Other geometries,  like \emph{e.g.} $\alpha =  \pi/2$, seem to
be free of corrections (see  Fig.~\ref{ap:0.5000} in the appendix).

\section{Universality of the locking relation}
\label{sec:fBM}

We start this section by noting that the locking relation
Eq.(\ref{eq:relation}), is satisfied by Markovian diffusive dynamics.
In normal diffusion, the MSD scales linearly in time and thus,
$\eta=1$. Moreover, the Sparre-Andersen theorem holds and therefore
$\beta=3/2$. Finally, it is easy to check that the transmission decays
as $T(L)\sim 1/L$ (see \emph{e.g.} \cite{weiss1982}).  In this section
we show that the validity of Eq.~\ref{eq:relation} goes beyond the
deterministic transport in polygonal billiards and diffusion.

Fractional Brownian motion (fBM) is a class of random walks
characterized by long-range correlated steps. fBm is one of the most
widespread Gaussian models for anomalous diffusion \cite{eliazar2024}.

Let us consider a fractional Brownian motion $B_t$ in one dimension
with $B_0=0$ and zero expectation value for all $t \in [0, T]$, where
$T$ is the total observation time.  The fBM is a continuous-time
Gaussian process defined through its covariance function given by
\cite{mandelbrot1968}:
\begin{eqnarray}\label{eq:cov}
{\rm Cov}\left(B_t,B_s\right) &=& \mathbb{E}\{\left(B_t - 
\mathbb{E}\{B_t\}\right)\left(B_s - \mathbb{E}\{B_s\}\right)\}  \nonumber\\
&=& \frac{K}{2} \left(t^{2 H} + s^{2 H} - |t - s|^{2 H}\right)\,,
\end{eqnarray}
where $K/2$ plays the role of a generalised diffusion coefficient and
the Hurst exponent $H\in(0,1)$. fBM is a generalization of Brownian
motion with stationary but correlated increments, yielding the process
non Markovian, except for $H=1/2$ corresponding to standrd Brownian
motion.

From Eq.(\ref{eq:cov}), the time dependence of the MSD for fBM is
algebraic, with exponent $\eta_{\mathrm{fBM}}=2H$.  The Hurst index
describes the raggedness of the resulting motion, with a higher value
leading to a smoother motion.  For $H < 1/2$ the increments of the
process are negatively correlated so that the fBM is subdiffusive.  On
the other hand, for $H > 1/2$ the increments of the process are
positively correlated and superdiffusive behavior occurs.

The transmission coefficient of fBM in a finite interval was derived
in Ref.~\cite{majumdar2010}.  Consider a stochastic process $\xi(x,t)$
in the finite interval $(0,L)$ with absorbing boundaries. Let $Q_L(x)$
be the probability that the process hits the boundary at $L$ before
hitting the boundary at $0$, given that the process started at
$\xi(0)=x\in(0,L)$. In Ref.~\cite{majumdar2010} it was shown that if
$\xi(x,t)$ is self-affine, namely if $\xi(x,t)=\xi(x/t^H)$ holds, then
$Q_L(x) = Q(x/L)$ and more importantly, that for $x/L \ll 1$,
$Q_L(x) \sim \left(x/L\right)^{\theta/H}$, where $H$ the Hurst index
and $\theta$ is the persistence exponent of fBM.

Noticing that for $x=0^+$, the probability $Q_L(x)$ is nothing but the
transmission coefficient $T(L)$ defined in section~\ref{sec:relation},
then, for fBM we have $T_{\mathrm{fBM}}(L) \sim
L^{-\theta/H}$. Furthermore, recalling that the persistence exponent
of fBM is $\theta= 1-H$ \cite{molchan1999,aurzada2011,bray2013}, we
obtain
\begin{equation}
T_{\mathrm{fBM}}(L) \sim L^{-(1-H)/H} \ .
\end{equation}
Therefore, the transmission coefficient decreases algebraically with
the length of the interval, with exponent $\nu = (1-H)/H$.

Finally, in Ref.~\cite{ding1995} it was shown that for fBM in the
finite interval $(0,L)$, the probability density of the first return
time to the origin is $P_0(\tau) \sim \tau^{H-2}$, and therefore
$\beta=H-2$.

\begin{figure}[!t]
\centering
   \includegraphics[width=0.45\textwidth]{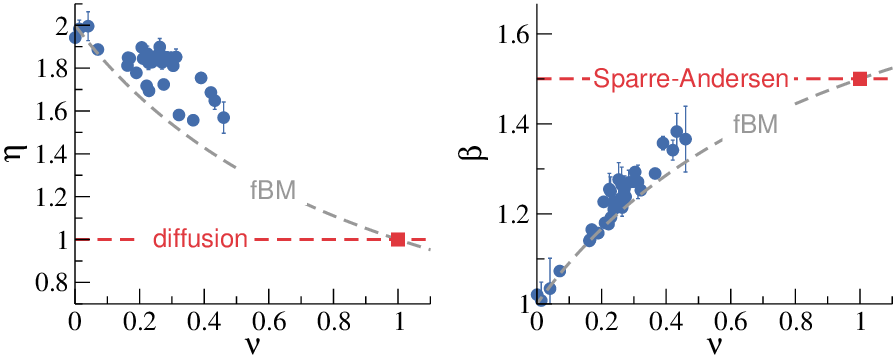}
   \caption{Sections of the parameter space of exponents $\eta$,
     $\beta$, and $\nu$.  Blue circles correspond to the values of the
     exponents obtained numerically for different polygonal
     channels. The gray dashed curve stands for fBM and the red square
     to the corresponding values of Brownian motion.  Red dashed lines
     shows indicate the value for normal diffusion $\eta=1$ (left
     panel) and for the Sparre-Andersen class $\beta=3/2$ (right
     panel).}
 \label{fig:exponents}
\end{figure}

The exponents for fBM in the finite interval are collected in
table~\ref{tab:exponents} and do satisfy the relation
\ref{eq:relation}. Thus, it appears that fBM and the deterministic
transport in polygons share their non Markovian character and to some
extent, the correlated increments.  In Fig.~\ref{fig:exponents} we
show the parameter space of the scaling exponents obtained numerically
for the different polygonal channels. In the figure, the gray dashed
curve stand for fBM and the red square to the corresponding values of
Brownian motion.

The interdependency between the exponents $\eta$, $\beta$, and $\nu$,
as established by Eq.~\ref{eq:relation}, shows that particle diffusion
as described by the MSD, the statistics of the first return times, and
the properties of the transmission, are intimately related. The fact
that this relation holds for such diverse dynamics leads us to
conjecture that the locking relation \ref{eq:relation} is likely to be
valid for other Markovian and non Markovian processes with a finite
MSD.

\section{Conclusions}
\label{sec:concl}

In summary, we have shown the the statistics of the first return times
in polygonal channels of finite length satisfies a simple scaling that
depends on the algebraic decay of $P_0(\tau)$ itself, and of the
scaling exponent of the MSD, $\eta$. Finite polygons exhibit a second
algebraic decay described, up to an integer or half-integer constant,
by the first decay. We speculate that the peculiarities of the second
decay may point to some kind of geometric selection rule.

Furthermore, Eq.~\ref{eq:relation} shows that the asymptotic behavior
of the MSD, the times of return to the origin and the way transmission
decays with the system size, are interdependent processes. Even when
exponents obtained from fits to power-laws are to be taken with a
\emph{grain of salt}, Eq.~\ref{eq:relation} seems robust.  As a matter
of fact, the existence of the two algebraic decays in the $P_0(\tau)$
separated by a time that depends on the way trajectories explore the
system in terms of the MSD, strongly simplifies the calculation of the
various exponents that describe the system.  Consider, for example,
having the statistics of the first return times $P^{(1)}_0(\tau)$ for
a system of length $L_1$ and $P^{(2)}_0(\tau)$ for a system of length
$L_2 \ne L_1$. From the densities, the exponent $\beta$ can be
estimated from a fit to a power-law. Knowing that $P^{(1)}_0(\tau)$
and $P^{(2)}_0(\tau)$ collapse into a single function after the
rescaling of Eq.~\ref{eq:collapse}, we can perform a numerical
optimization that minimizes the square distance by varying
$\eta\in(1,2)$.  The optimal solution leads to the estimation
$\eta$. Finally, using $\beta$ and $\eta$ we estimate $\nu$ through
Eq.~\ref{eq:relation}. Therefore, the knowledge of the first return
times for two lengths suffices in principle, to determine the
asymptotic behavior of the MSD and of the $T(L)$.

This leads to the conclusion that processes that are studied
independently are actually intimately related to each other, and that
their asymptotic properties satisfy a locking relation. We have also
shown that, this relation holds for the fractional Brownian motion
model of anomalous transport as well.  This leads us to conjecture
that the locking relation Eq.~\ref{eq:relation} is likely to be valid
for many Markovian or non Markovian processes exhibiting anomalous
transport possessing a finite MSD. This deserves further
investigation.

\section*{Acknowledgements}

We thank fruitful discussions with Eli Barkai and Rafaella Burioni.
CMM acknowledges financial support from the Spanish Government grant
PID2021-127795NB-I00 from MCIN/AEI/10.13039/501100011033 and FEDER,
UE.

\bibliography{polygon}

\newpage
\onecolumngrid
\appendix

\section{Exploring rational and irrational polygons}
\label{app}

To support the generality of our results over different polygonal
channels, in this appendix we show the numerical results for the
collapse of the PDF of the first return times to the origin
$P_0(\tau)$, as given by Eq.~\ref{eq:collapse}, and the decay of the
transmission coefficient as a function of the channel length, for
different polygonal cells.

\begin{figure}[!t]
\centering
   \includegraphics[width=1\textwidth]{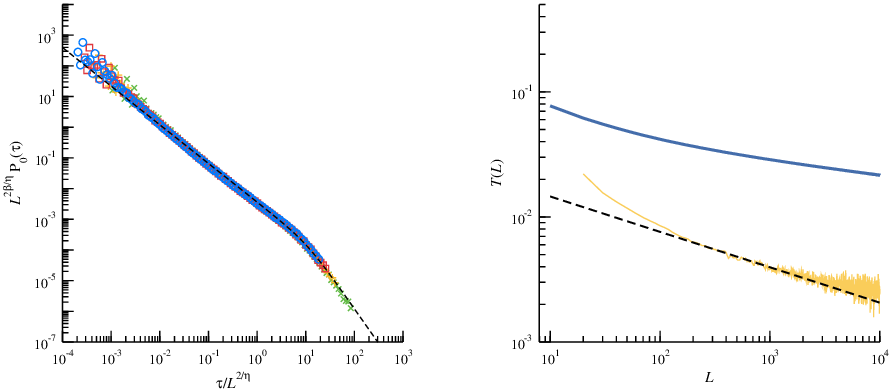}
   \caption{Panel (\emph{a}) Probability density of the first return
     times $P_0(\tau)$ scaled as in Eq.~\ref{eq:collapse} for a
     channel with $\alpha=\pi/4$, critical horizon, and length
     $L=2500$ (green crosses), $L=5000$ (yellow pluses), $L=7500$ (red
     squares), and $L=10000$ (blue circles). The dashed curve
     corresponds to Eq.~\ref{eq:ratfit} for $A=0.04$, $c=11$ and
     $\zeta=2$. Panel (\emph{b}) Transmission as a function of the
     system size $T(L)$ (dark blue). The yellow curve corresponds
     $-L T^\prime(L)$ and the dashed curve to $\sim L^{-\nu}$ with
     $\nu$ obtained from Eq.~\ref{eq:relation}.}
    \label{ap:0.2500}
\end{figure}

\begin{figure}[!t]
\centering
   \includegraphics[width=1\textwidth]{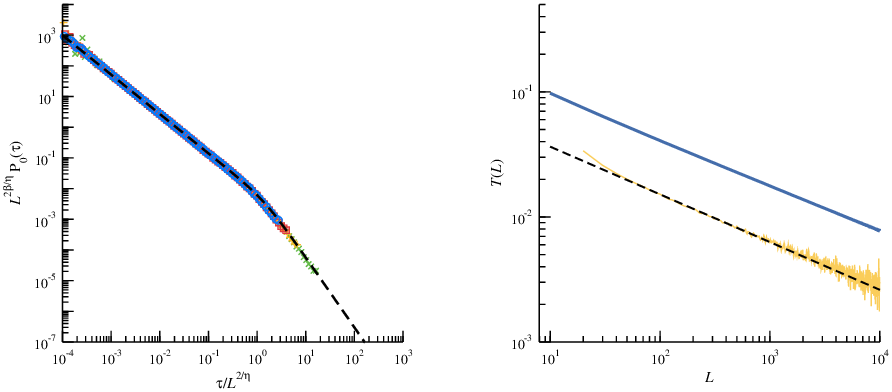}
   \caption{Panel (\emph{a}) Probability density of the first return
     times $P_0(\tau)$ scaled as in Eq.~\ref{eq:collapse} for a
     channel with $\alpha=\pi/3$, critical horizon, and length
     $L=2500$ (green crosses), $L=5000$ (yellow pluses), $L=7500$ (red
     squares), and $L=10000$ (blue circles). The dashed curve
     corresponds to Eq.~\ref{eq:ratfit} for $A=0.011$, $c=1.5$ and
     $\zeta=2$. Panel (\emph{b}) Transmission as a function of the
     system size $T(L)$ (dark blue). The yellow curve corresponds
     $-L T^\prime(L)$ and the dashed curve to $\sim L^{-\nu}$ with
     $\nu$ obtained from Eq.~\ref{eq:relation}.}
    \label{ap:0.3333}
\end{figure}

\begin{figure}[!t]
\centering
   \includegraphics[width=1\textwidth]{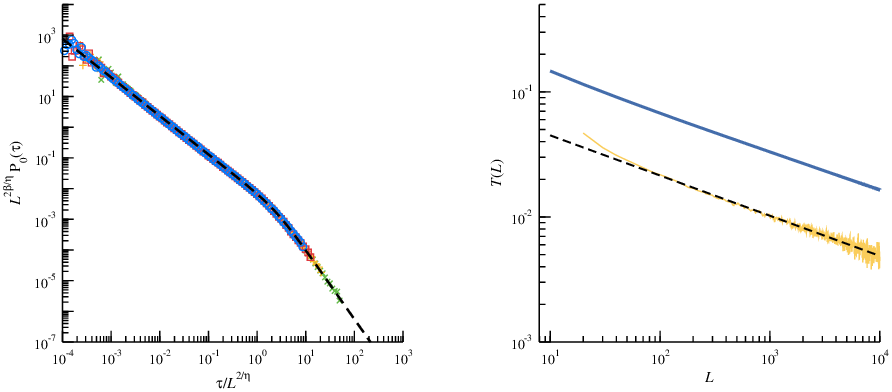}
   \caption{Panel (\emph{a}) Probability density of the first return
     times $P_0(\tau)$ scaled as in Eq.~\ref{eq:collapse} for a
     channel with $\alpha=\pi/2$, critical horizon, and length
     $L=2500$ (green crosses), $L=5000$ (yellow pluses), $L=7500$ (red
     squares), and $L=10000$ (blue circles). The dashed curve
     corresponds to Eq.~\ref{eq:ratfit} for $A=0.018$, $c=2.5$ and
     $\zeta=2$. Panel (\emph{b}) Transmission as a function of the
     system size $T(L)$ (dark blue). The yellow curve corresponds
     $-L T^\prime(L)$ and the dashed curve to $\sim L^{-\nu}$ with
     $\nu$ obtained from Eq.~\ref{eq:relation}.}
    \label{ap:0.5000}
\end{figure}

\begin{figure}[!t]
\centering
   \includegraphics[width=1\textwidth]{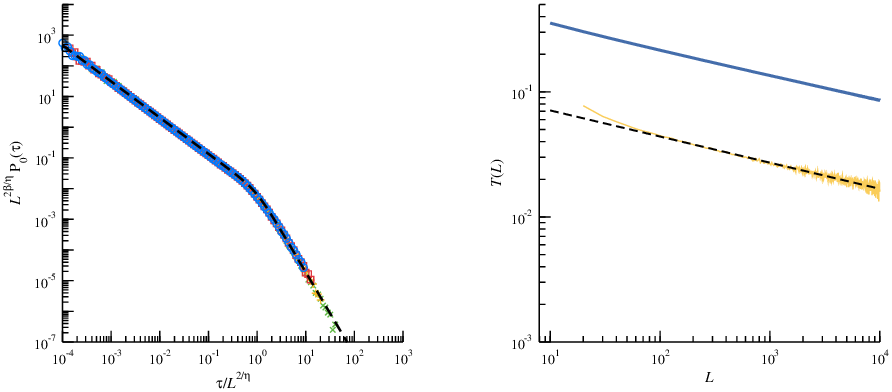}
   \caption{Panel (\emph{a}) Probability density of the first return
     times $P_0(\tau)$ scaled as in Eq.~\ref{eq:collapse} for a
     channel with $\alpha=2\pi/3$, critical horizon, and length
     $L=2500$ (green crosses), $L=5000$ (yellow pluses), $L=7500$ (red
     squares), and $L=10000$ (blue circles). The dashed curve
     corresponds to Eq.~\ref{eq:ratfit} for $A=0.0095$, $c=1$ and
     $\zeta=3$. Panel (\emph{b}) Transmission as a function of the
     system size $T(L)$ (dark blue). The yellow curve corresponds
     $-L T^\prime(L)$ and the dashed curve to $\sim L^{-\nu}$ with
     $\nu$ obtained from Eq.~\ref{eq:relation}.}
    \label{ap:0.6666}
\end{figure}

\begin{figure}[!t]
\centering
   \includegraphics[width=1\textwidth]{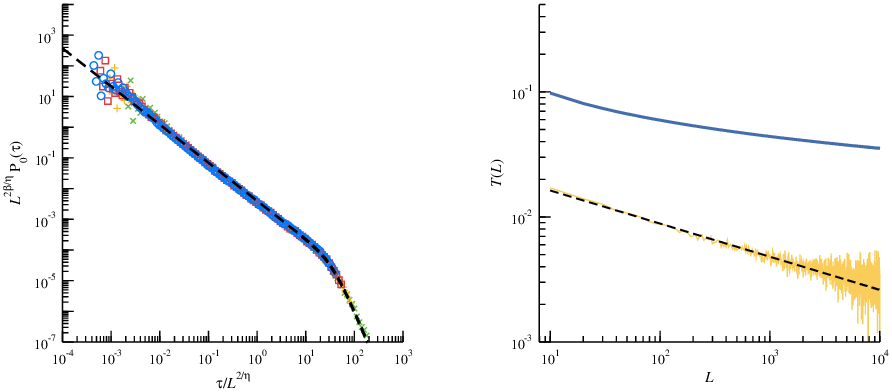}
   \caption{Panel (\emph{a}) Probability density of the first return
     times $P_0(\tau)$ scaled as in Eq.~\ref{eq:collapse} for a
     channel with irrational angle $\alpha=\xi \pi/4$, critical horizon, and length $L=2500$ (green crosses), $L=5000$ (yellow
     pluses), $L=7500$ (red squares), and $L=10000$ (blue circles). The dashed curve corresponds
     to Eq.~\ref{eq:irrfit} for $A=31$, $c=36$ and $\zeta=5/2$. Panel
   (\emph{b}) Transmission as a function of the system size $T(L)$
   (dark blue). The yellow curve corresponds $-L T^\prime(L)$ and the
     dashed curve to $\sim L^{-\nu}$ with $\nu$ obtained from Eq.~\ref{eq:relation}.}
    \label{ap:0.2500-irr}
\end{figure}

\begin{figure}[!t]
\centering
   \includegraphics[width=1\textwidth]{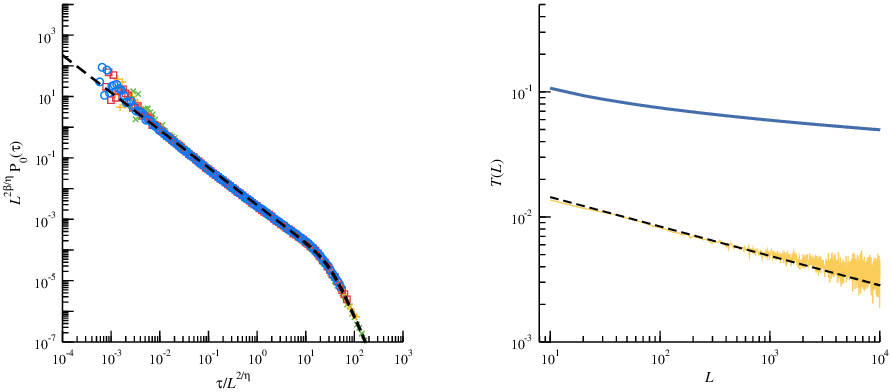}
   \caption{Panel (\emph{a}) Probability density of the first return
     times $P_0(\tau)$ scaled as in Eq.~\ref{eq:collapse} for a
     channel with irrational angle $\alpha=\xi \pi/3$, critical horizon, and length $L=2500$ (green crosses), $L=5000$ (yellow
     pluses), $L=7500$ (red squares), and $L=10000$ (blue circles). The dashed curve corresponds
     to Eq.~\ref{eq:irrfit} for $A=19.6$, $c=34$ and $\zeta=5/2$. Panel
   (\emph{b}) Transmission as a function of the system size $T(L)$
   (dark blue). The yellow curve corresponds $-L T^\prime(L)$ and the
     dashed curve to $\sim L^{-\nu}$ with $\nu$ obtained from Eq.~\ref{eq:relation}.}
    \label{ap:0.3333-irr}
\end{figure}

\begin{figure}[!t]
\centering
   \includegraphics[width=1\textwidth]{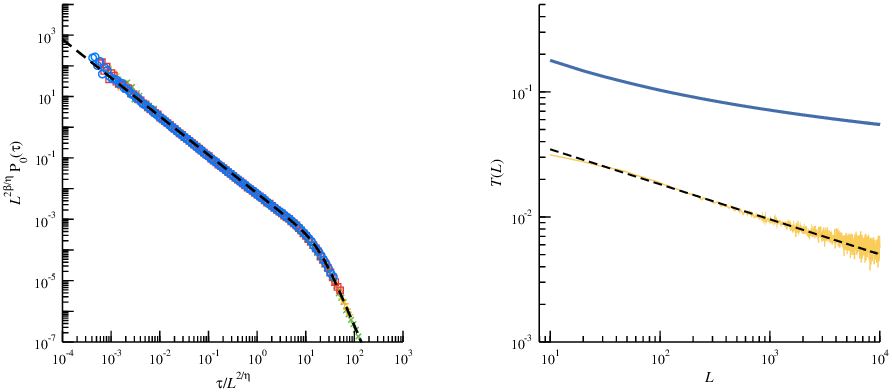}
   \caption{Panel (\emph{a}) Probability density of the first return
     times $P_0(\tau)$ scaled as in Eq.~\ref{eq:collapse} for a
     channel with irrational angle $\alpha=\xi \pi/2$, critical horizon, and length $L=2500$ (green crosses), $L=5000$ (yellow
     pluses), $L=7500$ (red squares), and $L=10000$ (blue circles). The dashed curve corresponds
     to Eq.~\ref{eq:irrfit} for $A=11$, $c=19$ and $\zeta=5/2$. Panel
   (\emph{b}) Transmission as a function of the system size $T(L)$
   (dark blue). The yellow curve corresponds $-L T^\prime(L)$ and the
     dashed curve to $\sim L^{-\nu}$ with $\nu$ obtained from Eq.~\ref{eq:relation}.}
    \label{ap:0.5000-irr}
\end{figure}

\begin{figure}[!t]
\centering
   \includegraphics[width=1\textwidth]{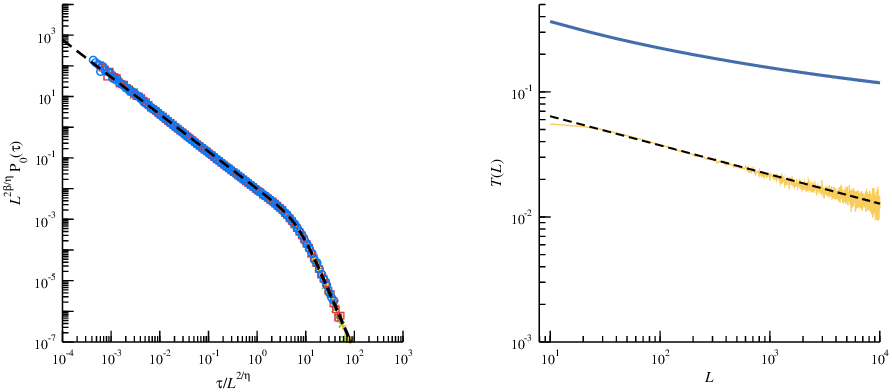}
   \caption{Panel (\emph{a}) Probability density of the first return
     times $P_0(\tau)$ scaled as in Eq.~\ref{eq:collapse} for a
     channel with irrational angle $\alpha=\xi 2\pi/3$, critical horizon, and length $L=2500$ (green crosses), $L=5000$ (yellow
     pluses), $L=7500$ (red squares), and $L=10000$ (blue circles). The dashed curve corresponds
     to Eq.~\ref{eq:irrfit} for $A=1.4$, $c=7.26$ and $\zeta=5/2$. Panel
   (\emph{b}) Transmission as a function of the system size $T(L)$
   (dark blue). The yellow curve corresponds $-L T^\prime(L)$ and the
     dashed curve to $\sim L^{-\nu}$ with $\nu$ obtained from Eq.~\ref{eq:relation}.}
    \label{ap:0.5000-irr}
\end{figure}

\end{document}